\documentclass{nature_with_images}

\usepackage{amsmath,bbm,amssymb}
\usepackage[squaren]{SIunits}
\usepackage[usenames,dvipsnames]{color}
\usepackage[pdftex]{graphicx}
\usepackage{hyperref}
\usepackage[normalem]{ulem}

\title{Certified quantum non-demolition measurement of a macroscopic material system}

\author{R.~J.~Sewell$^{1,*}$, M.~Napolitano$^1$, N.~Behbood$^1$, G.~Colangelo$^1$ \& M.~W.~Mitchell$^{1,2}$}

\definecolor{mygreen}{rgb}{0,0.5,0} 
\definecolor{myblue}{rgb}{0,0,0.75} 
\definecolor{mymagenta}{cmyk}{0,1,0,0.12}

\newcommand{\dd}{{\rm D}_2}

\newcommand{\dJ}{\delta J_{\rm s}}

\newcommand{\var}[1]{\mathrm{var}(#1)}
\newcommand{\cov}[1]{{\mathrm{cov}(#1})}
\newcommand{\dvar}[1]{\widetilde{\mathrm{var}}(#1)}
\newcommand{\dcov}[1]{\widetilde{\mathrm{cov}}(#1)}
\newcommand{\ave}[1]{\ensuremath{\langle#1\rangle}}

\newcommand{\DX}{\Delta X}
\newcommand{\DXs}{\Delta X_{\rm s}^2}
\newcommand{\DXm}{\Delta X_{\rm m}^2}
\newcommand{\DXsm}{\Delta X_{\rm s|m}^2}
\newcommand{\Ts}{T_{\rm s}}
\newcommand{\Tm}{T_{\rm m}}

\newcommand{\HH}{\hat{H}}

\newcommand{\SSS}{\hat{S}}
\newcommand{\Sx}{\SSS_{\rm x}}
\newcommand{\Sy}{\SSS_{\rm y}}
\newcommand{\Sz}{\SSS_{\rm z}}

\newcommand{\JJ}{\hat{J}}
\newcommand{\Jx}{\JJ_{\rm x}}

\newcommand{\Jz}{\JJ_{\rm z}}
\newcommand{\bJ}{\mathbf{\JJ}}

\newcommand{\jj}{\hat{j}}
\newcommand{\bj}{\mathbf{\jj}}

\newcommand{\ph}{\hat{\phi}}
\newcommand{\vph}{\hat{\varphi}}
\newcommand{\phA}{\ph_1}
\newcommand{\phB}{\ph_2}
\newcommand{\phC}{\ph_3}
\newcommand{\phABC}{\ph_{\{1,2,3\}}}
\newcommand{\phN}{\ph_{\rm RO}}

\newcommand{\XN}{X_{\rm RO}}
\newcommand{\YN}{Y_{\rm RO}}

\newcommand{\NA}{N_{\rm A}}
\newcommand{\NAtot}{\NA^{\rm (total)}}
\newcommand{\NL}{N_{\rm L}}

\newcommand{\rA}{r_{\rm A}}

\newcommand{\op}{\hat{O}}
\newcommand{\supin}{^{({\rm in})}}
\newcommand{\supout}{^{({\rm out})}}

\newcommand{\SyOut}{\Sy\supout}
\newcommand{\SyIn}{\Sy\supin}
\newcommand{\SxIn}{\Sx\supin}
\newcommand{\JzOut}{\Jz\supout}
\newcommand{\JzIn}{\Jz\supin}

\newcommand{\Jref}{J_0}

\newcommand{\opOut}{\op\supout}
\newcommand{\opIn}{\op\supin}
\newcommand{\vphOut}{\vph\supout}
\newcommand{\vphIn}{\vph\supin}
\newcommand{\phOut}{\ph\supout}

\newcommand{\ket}[1]{\left|#1\right>}

\begin{document}

\maketitle

\begin{affiliations}
 \item ICFO-Institut de Ciencies Fotoniques, Av. Carl Friedrich Gauss, 3, 08860 Castelldefels, Barcelona, Spain.
 \item ICREA-Instituci\'{o} Catalana de Recerca i Estudis Avan\c{c}ats, 08015 Barcelona, Spain.
\end{affiliations}

\begin{abstract}
Quantum non--demolition (QND) measurements improve sensitivity by evading measurement back-action.\cite{Braginsky1980} 
The technique was first proposed to detect mechanical oscillations in gravity wave detectors,\cite{Braginsky1975} and demonstrated in the measurement of optical fields,\cite{Roch1992,Grangier1998} leading to the development of rigorous criteria to distinguish QND from similar non-classical measurements.\cite{Grangier1998} 
Recent QND measurements of macroscopic material systems such as atomic ensembles,\cite{Kuzmich2000,Takano2009,Appel2009,Schleier-Smith2010a,Chen2011,Sewell2012} and mechanical oscillators,\cite{Thompson2008,Hertzberg2010,Vanner2012} show some QND features, but not full QND character.
Here we demonstrate certified QND measurement of the collective spin of an atomic ensemble.
We observe quantum state preparation (QSP) and information--damage trade--off (IDT) beyond their classical limits by seven and twelve standard deviations, respectively.  
Our techniques complement recent work with microscopic systems,\cite{Ralph2006,Lupascu2007,Neumann2010} and can be used for quantum metrology\cite{Takano2009,Appel2009,Schleier-Smith2010a,Chen2011,Sewell2012,Inoue2013} and memory,\cite{Jensen2011} the preparation\cite{Massar2003} and detection\cite{Dubost2012} of non--gaussian states, and proposed quantum simulation\cite{Eckert2007,Eckert2008,Hauke2013} and information\cite{Takano2008,Marek2010} protocols.
They should enable QND measurements of dynamical quantum variables\cite{Tsang2011,Eckert2007,Eckert2008} and the realization of QND-based quantum information protocols.\cite{Massar2003,Takano2008,Marek2010}
\end{abstract}

In a QND measurement, \emph{meter} ($X_{\rm m}$) and \emph{system} ($X_{\rm s}$) variables interact via an appropriate Hamiltonian and become entangled.  
Direct measurement of $X_{\rm m}$ then provides indirect information about $X_{\rm s}$ without destroying the system or altering $X_{\rm s}$. 
In the formulation of Roch {\em et al.},\cite{Roch1992} QND measurement of continuous variables, as used to describe systems with more than a few particles, is quantified by three figures of merit:  $\DX_{\rm m}^2$ describes the measurement noise referred to the input, $\DX_{\rm s}^2$ describes the variance in $X_{\rm s}$ added by the QND interaction, and $\DXsm$ describes the post--measurement conditional variance, i.e., the uncertainty in $X_{\rm s}$ given the measurement outcome.  
Two non--classicality criteria must be met to certify QND measurement:\cite{Roch1992,Grangier1998} $\DXsm<1$ describes a non--classical {\em quantum state preparation} (QSP) capability, while $\DXs\DXm<1$ describes a non--classical {\em information--damage tradeoff} (IDT). 
This latter inequality is usually expressed in terms of transfer coefficients $\Ts{\equiv} 1/(1+\DXs)$ and $\Tm{\equiv} 1/(1+\DXm)$ as $\Ts+\Tm>1$.  
Similar criteria have been developed for discrete--variable systems such as qubits,\cite{Ralph2006,Lupascu2007,Neumann2010} but are beyond the scope of this manuscript.
Throughout, the unit of noise is the standard quantum limit of the system, in our case the spin projection noise.  
Note that some non-QND operations such as filtering and optimal cloning can satisfy one or the other criterion.\cite{Grangier1998}

The QSP property describes the ability to generate quantum correlations between meter and {\em output} signal variables, i.e., at the end--point of the QND interaction.
This generates (conditional) squeezing\cite{Appel2009,Schleier-Smith2010a,Chen2011,Sewell2012,Vanner2012,Inoue2013} a resource for metrology or quantum information, but does not guarantee that the system variable was well measured.  
In the extreme, the signal and meter could finish in a perfectly correlated state which is completely unrelated to the input signal.  
In contrast, the IDT property involves the ability to correlate the meter to the {\em input} system variable, i.e., at the {start} of the QND interaction.  
This is valuable for any precise measurement, but does not imply measurement--induced squeezing.
For example, the measurement could faithfully copy the signal onto the meter, $\DXm\approx1$, before adding two units of quantum noise to the signal, $\DXs=2$.  
This satisfies IDT, as $\Ts+\Tm\approx\tfrac{4}{3}>1$, but it leaves the system in an extra--noisy state.
Satisfying both QND criteria implies the generation of quantum correlations between the meter and both the input {\it and} output state of the system variable.
This is important in metrological applications involving monitoring dynamical variables, such as in quantum waveform estimation\cite{Tsang2011} or the study of how quantum correlations in degenerate quantum gases evolve.\cite{Eckert2007,Eckert2008}
Repeated QND measurements of the same input state that satisfy both criteria are similarly required in various continuous-variable quantum information applications, for example, proposals for generating non--gaussian states,\cite{Massar2003} or quantum information processing protocols.\cite{Takano2008,Marek2010}
We note that QND measurement of dynamical variables has recently been placed in a more general theoretical framework in Ref.\cite{Tsang2012}

In optics, direct measurement of $X_{\rm s}$ and $X_{\rm m}$ can be compared against the QND criteria. 
With the macroscopic material systems used to date, $X_{\rm s}$ is not directly measurable with quantum--limited precision.  
Nevertheless, the statistics of $X_{\rm m}$ from two repeated QND measurements, e.g., conditional variances have been used to demonstrate QSP.\cite{Appel2009,Schleier-Smith2010a,Chen2011,Sewell2012}
However, the statistics of only two pulses are insufficient to verify the non--classical IDT criterion.
We note also that these experiments required independent measurements of the system coherence before and after the QND measurement to characterize damage done to the initial state and establish the reference quantum noise for verifying QSP.
Such measurements are not possible in many proposed QND applications, for example with unpolarized atomic ensembles.\cite{Eckert2007,Eckert2008,Toth2010,Hauke2013}
As shown in reference,\cite{Mitchell2012} statistics of \emph{three} successive QND measurements are sufficient both to find $\Ts$ and $\Tm$, and to quantify damage to the measured variable, and thus verify both the QSP and the IDT criteria.

\begin{figure}
	\centering
	\includegraphics[width=0.5\columnwidth]{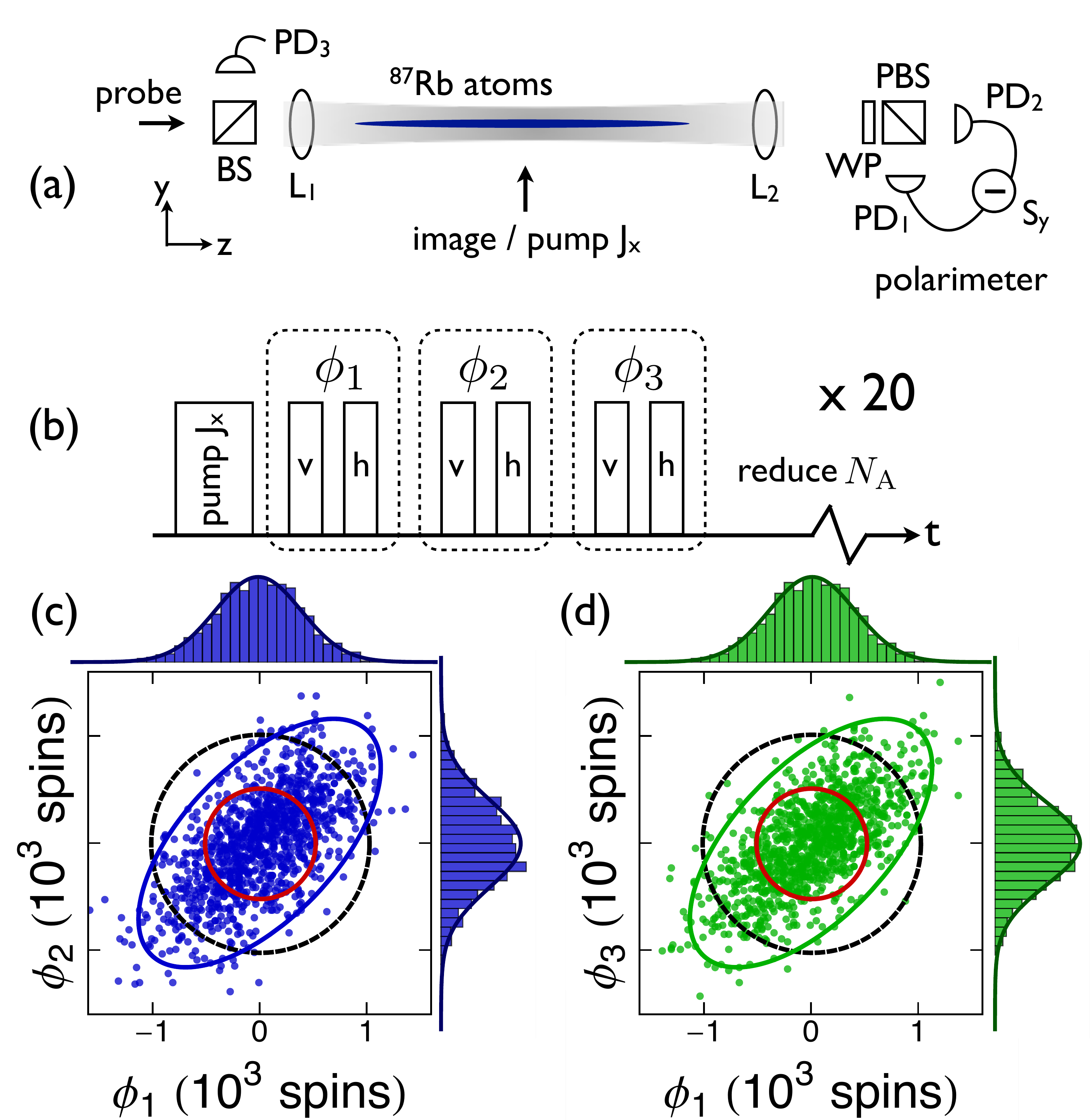}%
	\caption{
	{\bf (a)} Experimental geometry.  PD: photodiode; L: lens; WP: waveplate; BS: beam-splitter; PBS: polarizing beam-splitter.  {\bf (b)} 
	Measurement pulse sequence. 
	See text for details.  
	Also shown are the correlations between {\bf (c)} the first two QND measurements, and {\bf (d)} the first and third QND measurements, with $\NA=8.5\times10^5$ atoms.  
	The black dashed circles indicate the projection noise of the input coherent spin state, and the red solid circles indicate the measurement read-out noise (due to light shot noise).  
	The ratio of these gives the signal-to-noise of the QND measurement, a measure of the information transfer between the atomic system and meter variable.  
	The difference between the variance of consecutive measurements gives information about the noise introduced into the system variable by the QND measurement.  
	Damage to the system variable $\Jz$ is quantified by comparing the covariance between the first and third measurement to that between the first and second measurement.
	\label{fig:setup}}
\end{figure}

Here, we demonstrate certified QND of atomic spins via paramagnetic Faraday rotation in a quantum atom--light interface\cite{Hammerer2004}.
In our apparatus, described in detail in Ref.,\cite{Kubasik2009} and illustrated in Fig.~\ref{fig:setup}(a), we work with an ensemble of $f=1$ atoms held in an optical dipole trap and interacting with \unit{\micro\second} pulses of near--resonant light propagating along the $z$-axis.
The interaction between the atoms and each pulse of light is characterized by an effective Hamiltonian
\begin{equation}
	\tau\HH =  \kappa \Sz\Jz 
	\label{eqn:H_full}
\end{equation}
which describes an QND measurement of $\Jz$ via paramagnetic Faraday rotation: Pulses of light with an input polarization $\Sx$ and pulse duration $\tau$ experience a polarization rotation $\vphOut=\vphIn+\kappa\JzIn$ proportional to the collective atomic spin, leaving the spin variable unchanged, $\JzOut=\JzIn$ (see Methods).
For multi--level alkali atoms, this effective Hamiltonian can be synthesized using multicolor or dynamical--decoupling probing techniques.\cite{Appel2009,Koschorreck2010b}

For convenience, we define a scaled rotation angle $\ph\equiv\vph/\kappa$, so that $\phOut=\phN+\JzIn$, where $\phN = \SyIn/(\kappa\SxIn)$ is the (scaled) input polarization angle.
In our experiment the coupling constant $\kappa$ is calibrated from independent measurements (see Methods), but it can also be extracted from the noise scaling of a known input state.\cite{Koschorreck2010a}
Fluctuations of $\phN$ give the read--out noise $\var{\phN}$, which is directly observable by measuring without atoms in the trap.

In order to quantify the QND measurement variances we make three consecutive measurements $\phABC$ of $\Jz$ (see Fig.~\ref{fig:setup}(b)).
The conditional noise reduction is quantified using the first two measurements: $\var{\Jz|\phA}\equiv\var{\phA-\chi\phB}-\var{\phN}$, where $\chi\equiv\cov{\phA,\phB}/\var{\phA}$.\cite{Sewell2012} 
The normalized conditional variance is then:\cite{Mitchell2012}
\begin{equation}
	\DXsm\equiv\frac{\var{\Jz|\phA}}{\rA \Jref}
	\label{eqn:conditional_noise}
\end{equation}
where $\Jref\equiv\ave{\Jx}/2=\NA/4$ is the projection--noise of the input atomic state, established from an independent measurement of the atom number $\NA$, and $\rA$ is the fraction of atoms that remain in the input state after the interaction.

To quantify the damage $\rA$ to the $\Jz$ variable due to the QND measurement without resorting to auxiliary measurements requires a comparison of the correlations among all three measurements (see Fig.~\ref{fig:setup}(c) \& (d)): In Ref.\cite{Mitchell2012} it is shown that $\rA\equiv\dcov{\phA,\phC}/\dcov{\phA,\phB}$, where we introduce the notation $\dvar{X}\equiv\var{X}-\var{\XN}$ and $\dcov{X,Y}\equiv\cov{X,Y}-\cov{\XN,\YN}$ for variances and covariances. 

The normalized meter and system variances can be written:\cite{Mitchell2012}
\begin{align}
	\DXm&\equiv\frac{\var{\phA}-\Jref}{\Jref}	\\
	\DXs&\equiv\frac{\dvar{\phB}-\dvar{\phA}}{\rA \Jref}
	\label{eqn:DeltaX}
\end{align}
and can be similarly quantified from the statistics of the three successive measurements.
The meter variance $\DXm$ is a measure of the information transfer between the atomic system and meter variable: $\var{\phA}$ includes quantum noise from the system variable (atomic projection noise), and a contribution (light shot--noise) from the meter variable that should be small to ensure that the system variable is measured with good precision.
The system variance $\DXs$ is a measure of the perturbation to the system variable due to the QND measurement.

We gain insight into the expected behaviour with a simple model of an ideal QND measurement.  
As described in Refs.\cite{Hammerer2004,Echaniz2005} we expect a conditional noise reduction by a factor $1/(1+d_0\eta)+2\eta$, where $d_0=(\sigma_0/A)\NA$ is the on--resonance optical depth of the atomic ensemble with $\sigma_0$ the on--resonance scattering cross--section and $A$ an effective interaction area, and $\eta$ is the probability that any given atom suffers decoherence due to spontaneous scattering from the probe beam.  
Note that $d_0\eta = \kappa^2\NA\NL/2$ is the ratio of atomic projection noise $\var{J_z}$ to readout noise $\var{\phN}$ at the standard quantum--limit, i.e., it is the signal--to--noise ratio when measuring a spin coherent state.

\begin{figure}
	\centering
	\includegraphics[width=0.75\columnwidth]{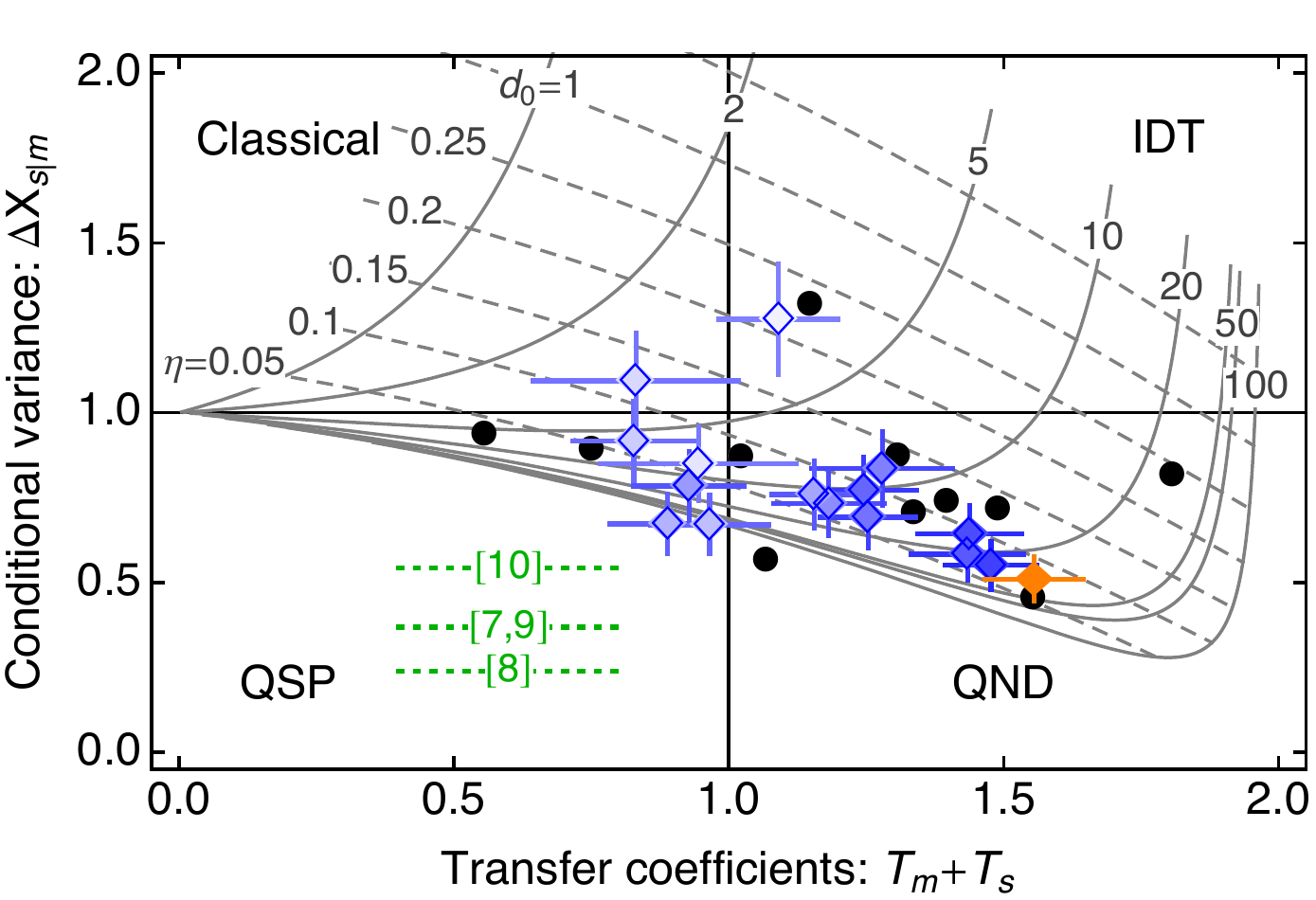}%
	\caption{
	Blue diamonds: Conditional variance and transfer coefficients quantified via three successive QND measurements.
	Shading represents a change in $\NA$ from $3.9\times10^4$ (light blue) to $8.5\times10^5$ (dark blue).  
	Error bars indicate $\pm1\sigma$ statistical errors.
	Orange diamond: best observed QND measurement with $\DXsm=0.64(5)$, $\Tm+\Ts=1.72(4)$ and $\NA = 8.5\times10^5$.
	Contours show the simple model described in Eqs.~\ref{eq:dxsm} and~\ref{eq:dxsdxm} using the measured disturbance parameter $\dJ=0.3(2)$; solid curves show contours of increasing on-resonant optical depth $d_0$, and dashed curves increasing $\eta$.  
	Black squares: QND measurements of optical fields reviewed in reference.\cite{Grangier1998}
	Green dotted lines: demonstrated QSP from spin squeezing results reported in Refs.\cite{Appel2009,Schleier-Smith2010a,Chen2011,Sewell2012}
	The transfer coefficients are unknown for these results.
	We do not include results from Refs.\cite{Kuzmich2000,Takano2009} for lack of an estimate of the damage to the measured state.
	Labels indicate the regions available to: classical measurements, quantum state preparation (QSP) without IDT, non-classical information-damage tradeoff (IDT) without QSP, and quantum non-demolition (QND).
	\label{fig:noiseInput}}
\end{figure}

The conditional variance can then be expressed as
\begin{equation}
	\DXsm=\frac{1}{(1+d_0\eta)(1-\eta)}+\frac{2\eta}{1-\eta}.
	\label{eq:dxsm}
\end{equation}
Similarly, the measurement noise referred to the input is just the inverse of the signal--to--noise ratio, $\DXm=1/d_0\eta$, and if we define $\dJ\equiv(\dvar{\phB}-\dvar{\phA})/\Jref$ (in units of the atomic projection noise) so that $\DXs=\dJ/(1-\eta)$, then we have
\begin{equation}
	\DXm\DXs=\frac{\dJ}{d_0\eta(1-\eta)}.
	\label{eq:dxsdxm}
\end{equation}
Eqs.~\ref{eq:dxsm} and~\ref{eq:dxsdxm} have different dependence on $d_0$ and $\eta$, with the result that some conditions satisfy QSP but not IDT and vice--versa.  
As shown in Fig. \ref{fig:noiseInput}, for sufficient $d_0$, low $\eta$ gives  QSP, high $\eta$ gives IDT, and intermediate $\eta$ can give both, i.e., QND.

In our experiment we trap between $3.9\times10^4$ and $8.5\times10^5$ atoms in a weakly focused single beam optical dipole trap.
We probe the atoms with \unit{2}{\micro\second} duration pulses of light propagating along the $z$-axis, with on average $2\times10^8$ photons per pulse and about \unit{1}{\giga\hertz} detuned from the $\dd$ line.
The trap geometry produces a large atom--light interaction for light propagating along the trap axis, characterized by the effective on--resonance optical depth $d_0$.
Dynamical--decoupling techniques allow us to make projection--noise--limited measurements of the spin--1 atoms using pairs of alternately h- and v-polarized pulses,\cite{Koschorreck2010b} with a demonstrated spin read--out noise of $(515\,{\rm spins})^2$.\cite{Koschorreck2010a}
From independent measurements, we estimate a maximum optical--depth $d_0=43.5$ and the probability of damage to any given atom's state due to scattering $\eta=0.093$.\cite{Sewell2012}

The results of the measurement as a function of increasing atom number $\NA$ are shown as blue diamonds in Fig.~\ref{fig:noiseInput}.  
A conditional variance $\DXsm<1$ indicates successful QSP and results in a spin--squeezed atomic state.\cite{Sewell2012}
With $\NA=8.5\times10^5$ atoms, we measure $\DXsm=0.64(5)$, with a fraction of atoms remaining in the initial state $\rA=0.76(4)$.  
The normalized meter and system variances with the same number of atoms were $\DXm=0.11(5)$ and $\DXs=0.23(1)$, giving $\Tm+\Ts=1.72(4) > 1$, demonstrating a non--classical IDT, thus fulfilling both criteria for certified QND measurement.
For comparison, with our parameters the simple model of Eqs.~\ref{eq:dxsm} and~\ref{eq:dxsdxm} gives a QSP parameter $\DXsm=0.42(2)$ and IDT parameters $\DXm=0.25(3)$ and $\DXs=0.3(2)$, or $\Tm+\Ts=1.6(1)$.

QND measurement techniques play increasingly important role in diverse applications, from quantum metrology\cite{Appel2009,Schleier-Smith2010a,Chen2011,Sewell2012,Inoue2013} and quantum memory,\cite{Jensen2011} to proposals for producing\cite{Massar2003} and detecting\cite{Dubost2012} non--gaussian states, continuous--variable quantum information processing protocols\cite{Takano2008,Marek2010} and in hybrid quantum devices.\cite{Hammerer2009}
The pulsed measurement techniques demonstrated here can be applied to QND measurement of any continuous variable material system, and complement similar criteria established for discrete variable experiments.\cite{Ralph2006,Lupascu2007,Neumann2010}
They may be particularly useful in experiments in which auxiliary measurement of the system coherence are not possible, such as applications requiring unpolarized atomic ensembles.\cite{Toth2010,Hauke2013}
We note also that pulsed QND measurement techniques have recently been used to demonstrate ponderomotive squeezing, measurement-induced cooling and quantum state tomography in nanomechanical oscillators.\cite{Vanner2012}
Verifying that both the QSP and IDT criteria of QND measurement are satisfied will be important in applications requiring multiple repeated measurements of the same system, such as in monitoring dynamical quantum variables\cite{Tsang2011,Tsang2012,Eckert2007,Eckert2008} and in some continuous-variable quantum information protocols.\cite{Massar2003,Takano2008,Marek2010}

\begin{methods}
\label{sec:methods}
\subsection{Atom-light interaction}
The atoms are described by collective spin $\bJ\equiv\sum_{n}^{}{\bj}^{(n)}$ where $\bj$ is a pseudo spin--1/2 operator on the $\ket{f=1,m_f = \pm 1}$ subspace and the sum runs over the $\NA$ atoms in the ensemble.  
The light pulse, with $N_L$ photons on average, is described by the Stokes operator $\SSS_i \equiv \tfrac{1}{2}(a^\dagger_L,a^\dagger_R) \sigma_i  (a_L,a_R)^T$ where $a_{L,R}$ are annihilation operators for the left-- and right--circular polarizations and $\sigma_i$ are the Pauli matrices.  
The input pulses are polarized with $\ave{\Sx} = N_L/2$.
To lowest order in the atom-light coupling constant $\kappa$ --- which depends on the trap and probe beam geometry, excited--state linewidth, laser detuning, and the hyperfine structure of the atom --- the interaction described by Eq.~\ref{eqn:H_full} produces a rotation of the state: $\opOut=\opIn-i\tau[\opIn,\HH]$.  
For an input atomic polarization $\ave{\Jx}=\NA/2$, corresponding to a coherent spin state (CSS), this imprints information about the system variable $\Jz$ onto the measurement variable $\Sy$ without changing $\Jz$
\begin{eqnarray}
	\SyOut&=&\SyIn+\kappa\SxIn\JzIn\\
	\JzOut&=&\JzIn
	\label{eqn:intput_output}
\end{eqnarray}
which describe a QND measurement of the collective atomic spin $\Jz$.
The parameter $\kappa\Sx$ parametrizes information transfer between the atoms and light, however increasing $\kappa\Sx$ also increases damage to the atomic state via spontaneous scattering.\cite{Hammerer2004}
\subsection{Measurement cycle}
In each experimental cycle we prepare a coherent spin state $\ave{\Jx}=\NA/2$ via optical pumping and make three successive QND measurements using a train of \unit{\micro\second} pulses of light with alternating h-- and v--polarization, at a detuning of \unit{600}{\mega\hertz} to the red of the $f=1 \rightarrow f'=0$ transition on the $\dd$ line, detected by a shot--noise--limited polarimeter.  
We synthesize the interaction described in Eq.~\ref{eqn:H_full} by combining the measurement results of consecutive pulses with orthogonal polarization.\cite{Koschorreck2010b,Sewell2012}
We vary the number of atoms, $\NA$, from $3.9\times10^4$ to $8.5\times10^5$ by briefly switching off the optical dipole trap for \unit{100}{\micro\second} after each measurement, which reduces the atom number by $ \sim15\,\% $, and repeating the sequence 20 times per trap loading cycle. 
At the end of each cycle the measurement is repeated without atoms in the trap.  To collect statistics, the entire cycle is repeated $\sim$ 1000 times.
\subsection{Calibration}
The coupling constant $\kappa=1.47\times10^{-7}$ radians per spin is calibrated against a measurement of the atom number made by absorption imaging.\cite{Koschorreck2010a}  
To account for the spatial variation in the coupling between the probe beam and the trapped atoms, we define an effective atom number such that the parametric Faraday rotation signal is proportional to the total number of atoms, and the expected variance of the measurement variable is $\var{\Jz}\equiv\NA/4$.\cite{Appel2009,Schleier-Smith2010a,Chen2011}  
For our trap and probe geometry $\NA=0.9\NAtot$.\cite{Sewell2012}
\end{methods}



\begin{addendum}
\item This work was supported by the Spanish MINECO under the project MAGO (Ref. FIS2011-23520) and by the European Research Council  under the project {AQUMET}, and by Fundaci\'{o} Privada CELLEX Barcelona.
\item[Correspondence] $^{*}$Correspondence and requests for materials should be addressed to R.~J.~Sewell~(email: robert.sewell@icfo.es).
\end{addendum}


\end{document}